\documentclass[reprint,twocolumn,showpacs,showkeys]{revtex4-1}

\usepackage{latexsym,graphicx,graphics}
\usepackage{appendix}
\usepackage{amsmath,amssymb,epsfig}
\usepackage{float}
\usepackage[inline]{enumitem}

\usepackage{color}

\definecolor{dark-green}{rgb}{0.18,0.55,0.2}                    
\definecolor{dark-blue}{rgb}{0.0,0.3,0.4}

\definecolor{dark-red}{rgb}{0.75,0.0,0.1}                    

\newcommand{\abs}[1]{\left| #1 \right|}

\newcommand{\bb}[1]{ \mbox{\boldmath$ #1$}}

\newcommand{\rv}{\bb r}

\newcommand{\rhov}{\bb{\rho}}
\newcommand{\rhovp}{\bb{\rho'}}

\newcommand{\Eq}[1]{Eq.~(\ref{#1})}
\newcommand{\Eqs}[2]{Eqs.~(\ref{#1})--(\ref{#2})}

\newcommand{\Gst}{G^{\mbox{\tiny st}}}
\newcommand{\Gstp}{G^{\mbox{\tiny st}(1)}}

\newcommand{\unit}[1]{\bb{\hat{#1}}}

\newcounter{denselistcounter}

\begin{document}

\title{Rest frame interference in rotating structures and metamaterials}

\author{Y.~Mazor and Ben Z. Steinberg}
\affiliation{School of EE, Tel-Aviv University}

\date{\today}

\begin{abstract}
Using the formulation of electrodynamics in rotating media, we put into explicit quantitative form the effect of rotation on interference and diffraction patterns as observed in the rotating medium's rest-frame. As a paradigm experiment we focus the interference generated by a linear array of sources in a homogeneous medium. The interference is distorted due to rotation; the maxima now follow curved trajectories. Unlike the classical Sagnac effect in which the rotation induced phase is independent of the refraction index $n$, here the maxima bending increases when $n$ decreases, suggesting that $\epsilon$-near-zero metamaterials can enhance optical gyroscopes and rotation-induced non-reciprocal devices. This result is counter intuitive as one may expect that a wave that travels faster would bend less. The apparent contradiction is clarified via the Minkowski momentum picture for a quasi-particle model of the interference that introduces the action of a Coriolis force, and by the Abraham picture of the wave-only momentum. our results may also shed light on the Abraham-Minkowski controversy as examined in non-inertial electrodynamics.
\end{abstract}
\maketitle

\noindent
{\bf\emph{Introduction}}
Non-reciprocal electrodynamics finds diverse applications in technology and engineering. It constitutes the basis for devices such as one-way waveguides, circulators, and isolators, that are important building blocks in functional electromagnetic and acoustic systems. Essentially three types of schemes have been suggested for breach of reciprocity \cite{Nonrec_review_Caloz}.
\begin{enumerate*}[label=(\roman*)]
    \item Magnetic biasing and the magneto-optic effect \cite{Rayleigh_magnetic,Pozar_book}.
    \item Medium nonlinearity \cite{Nonlin1_Hubner,Nonlin2_Stedman}.
    \item Spatio-temporal variation of the material properties or its rotation \cite{YuFan_2009,LiraLipson_2012,ZanjaniLukes_2013,SounasAlu_2013,SounasAlu_2014,FleuryAlu_2014, Mario_Sil_Rotation_2016}.
\end{enumerate*}
In all of these schemes, interest is focused on the electrodynamic phenomena as observed in the laboratory (inertial) Frame of Reference (FoR).

Non-reciprocity based on rotating media is of interest not only for the devices mentioned above; it plays a pivotal role in rotation sensors and optical gyroscopes. Here, interest is focused on the device's electrodynamics as observed in the rotating structure \emph{rest} FoR, $\mathcal{R}^\Omega$, that is non-inertial. This is motivated by the actual use of rotation sensors for independent navigation systems. Their operation is mainly based on the Sagnac effect - a manifestation of the non-reciprocity of a rotating medium as observed in its rest frame \cite{Post,FORS_Book,Gyro_Lefevre_2,Gyro_Lefevre}. Note that vacuum observed in a rotating FoR can be viewed as a ``rotating vacuum'' - a special case of a rotating medium. These applications have motivated studies of rotating Photonic Crystals \cite{BZS1, BZS2} and various guiding structures \cite{BZS3, Scheuer1, Scheuer2}, the dynamics associated with mode degeneracy in rotating cavities \cite{BZS2, Sunada} as well as the electrodynamics of rotating open cavities \cite{HuiCao1, HuiCao2}.
The non-reciprocity of rotating structures as observed in $\mathcal{R}^{\Omega}$ may find applications not only for optical gyroscopes. The analysis of rigidly rotating structures in $\mathcal{R}^\Omega$ offers significant advantages by the fact that the need to deal with moving boundaries is alleviated (albeit at the expense of more complicated constitutive relations \cite{Shiozawa}). Numerical solvers based on FDTD approach that make explicit use of this advantage were also developed \cite{HuiCaoNumer, Novitski}. Thus, even if the final results of interest are fields observed in the laboratory inertial FoR, one may be better off solving for the electrodynamic problem in $\mathcal{R}^\Omega$, and then apply the appropriate explicit field transformations \cite{VanBladel1,DeZutter2}. This general approach may therefore be suitable for the study of rigidly rotating complex structures such as metamaterials and metasurfaces, and may open the way for a new realm of applications for optical gyroscopes and other non-reciprocal devices.

In this work we present a preliminary study of the effect of slow rotation on the interference generated by a set of electrically small and equally excited scatterers embedded in a rigidly rotating material with constant angular velocity $\bb{\Omega}$. We explore the rotation footprint on the interference pattern observed at the structure's rest FoR $\mathcal{R}^{\Omega}$, and show that the Interference Maxima (IM) follow curved trajectories, whose curvature depends on $\Omega$. While this result by itself is intuitive, there are two new fundamental observations about the rotation footprint. First, unlike the traditional Sagnac effect which is medium-independent \cite{Post,Gyro_Lefevre_2,Gyro_Lefevre}, here the bending of the IM depends on the medium refraction index $n$. Second, the bending \emph{increases} as $n$ \emph{decreases}. The last observation is counter-intuitive: one would expect that if the wave travels faster for a given $\Omega$, its bending would decrease. We resolve the apparent contradiction by using, respectively, the Abraham and Minkowski pictures for the energy and momentum of the associated wave and quasi-particle interpretations of the interference pattern. Clearly, the fact that the rotation footprint on interference increases as $n$ decreases is a fundamentally new result that calls for novel applications of metamaterials and near-zero index materials in optical gyroscopes and other rotation based non-reciprocal devices.

\noindent
{\bf \emph{Formulation}}
We \emph{define} $\mathcal{R}_I$ as a static inertial FoR, in which the basic physical laws appear in their simple familiar form; the laws of \emph{Special Relativity} apply, space-time is flat (gravitation neglected) hence the system is \emph{Lorentzian} \cite{G_MTW}, and Maxwell’s equations take on their form in vacuum. A stationary material present in this system is represented here by the scalars $\epsilon(\rv),\mu(\rv)$.

Our interest is in an interference experiment in materials and structures that rotate rigidly at a constant angular velocity $\bb{\Omega}=\Omega\unit{z}$ with respect to $\mathcal{R}_I$. The spatial extent $d$ normal to $\unit{z}$ occupied by the material is finite, and the rotation is slow such that $\Omega d\ll c$. Since the system is Lorentzian, one may define this rotation with respect to $\mathcal{R}_I$ in the na\"{\i}ve way: follow a specific atom, measure the time interval $\Delta t$ until it accomplishes a complete round, then set $\Omega=2\pi/\Delta t$. We focus on the Electrodynamics (ED) as seen by an observer that is \emph{fixed} to the rotating material/structure. Hence we define the rotating FoR $\mathcal{R}^\Omega$. This is the \emph{appropriate reference frame} \cite{G_MTW}: it is at rest relative to all apparatus, sensors, etc., that are “bolted” to the material.
Here and henceforth, we always observe the ED as seen in the (always) appropriate FoR $\mathcal{R}^\Omega$, and the corresponding ED problem is termed as the ``$\mathcal{R}^{\Omega}$ problem''. The specific case of $\mathcal{R}^0$ and ``$\mathcal{R}^0$ problem" corresponds to $\Omega=0$, meaning that the material and the observer appear at rest in the inertial FoR $\mathcal{R}_I$, rendering $\mathcal{R}^0$ and $\mathcal{R}_I$ the same.

$\mathcal{R}^\Omega,\,\Omega\ne 0$ is non-inertial; an observer at rest there sees curved space-time. However, space-time in $\mathcal{R}^\Omega$ can be considered \emph{locally flat} (and Lorentzian) for distances $D$ satisfying \cite{G_MTW},
\begin{equation}\label{eqFlatness}
D\ll L=c^2/A,
\end{equation}
where $A$ is the acceleration of $\mathcal{R}^\Omega$. In our case $A=\Omega^2 D$, so the flatness criterion above yields \emph{precisely} the slow rotation condition mentioned above, $\Omega D\ll c$. Our work is limited to this domain of space-time flatness.

Our analysis is two-dimensional; the scatterers geometry and the electromagnetic excitation are invariant along $\unit{z}$ - the medium's rotation axis.
Vectors are written in bold letters, and a hat indicates a unit-vector. In $\mathcal{R}^{\Omega}$ the coordinates normal to the rotation axis $\unit{z}$ are denoted by $\bb{\rho}=\unit{x}x+\unit{y}y=\rho\unit{\rho}$. We assume that the materials in the corresponding $\mathcal{R}^0$ problem are fully characterized by scalar permittivity $\epsilon=\epsilon_0\epsilon_r$ and permeability $\mu=\mu_0\mu_r$. The corresponding speed of light in vacuum and refraction index for $\mathcal{R}^0$ problem are given by $c=1/\sqrt{\epsilon_0\mu_0}$ and $n=\sqrt{\epsilon_r\mu_r}$, respectively. The covariant formulation of electrodynamics of a slowly rotating medium, for $\mathcal{R}^\Omega$ problem, is given by the same set of Maxwell equations as for the stationary case except that the constitutive relations are modified to account for the rotation effect \cite{Shiozawa}. Our starting point is the Green's function theory developed in \cite{BZSG}. It has been shown that in 2D problems the electromagnetic field can be rigorously separated into completely decoupled TE and TM polarizations comprising of $(Hz,\, \bb{E}_t)$ and $(Ez,\, \bb{H}_t)$ fields, respectively. An exact Green's function expression has been derived, from which \emph{all} the field components can be uniquely extracted. Relevant results are summarized in the appendices including a formally exact representation in terms of cylindrical harmonics summation (e.g. \Eq{Aeq5a}). Under the slow rotation assumption $\max(\rho,\rho')\Omega/c\ll1$ this Green's function is given by
\begin{equation}\label{eq6}
G(\rhov,\rhovp) \simeq  \Gst(\rhov,\rhovp)e^{ik_0(\Omega/c)\unit{z}\cdot(\rhovp\times\rhov)},
\end{equation}
where $\Gst$ is the homogeneous medium Greens function for $\mathcal{R}^0$ problem,
$\Gst(\rhov,\rhovp )= \frac{i}{4}H_0^{(1)}(k_0n|\rhov-\rhovp|)$.

Up to a multiplication constant, the $\unit{z}$-directed fields $H_z$ and $E_z$ in the TE and TM polarizations, respectively, are given by this Green's function. Unlike the infinite summation over cylindrical harmonics, this approximation explicitly incorporates the physically meaningful entities of amplitude, phase, and optical path \cite{BZSG}.
We repeat for emphasis that $c$ is the speed of light in vacuum, for $\mathcal{R}^0$ problem. Thus, the rotation footprint is manifested only in the exponential term, and is \emph{independent} of the medium refraction index $n$.

\noindent
{\bf \emph{Reconstruction of the Sagnac effect}}
It has been shown in \cite{BZSG} that the full expression for $G$ (\Eq{Aeq5a}) contains the entire electrodynamic spectrum of a rotating medium, from which the Sagnac effect is obtained as a special case. The following simple analysis shows that the Sagnac phase shift is faithfully represented in \Eq{eq6}.

\begin{figure}[ht]
\centering\includegraphics[scale=0.4]{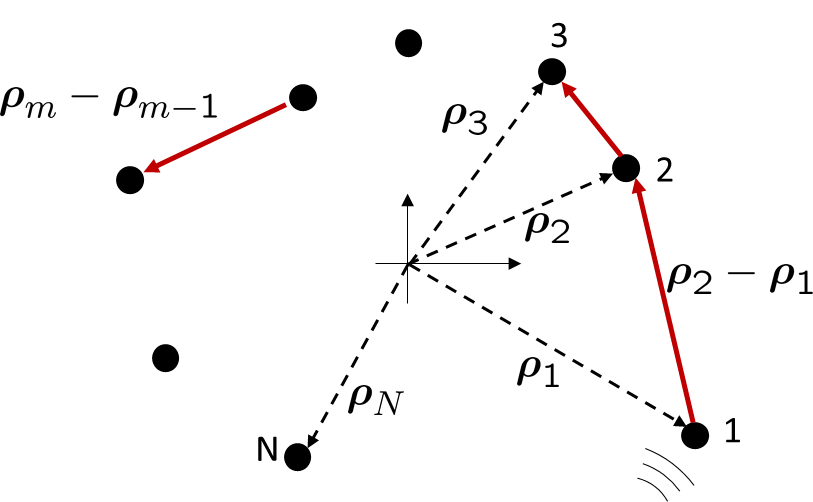}
\caption{A set of $N$ scatterers forming a Sagnac loop of scattering events. The structure rotates around the origin at angular velocity $\bb{\Omega}=\unit{z}\Omega$, and is observed at $\mathcal{R}^\Omega$.}
\label{fig1}
\end{figure}

Consider the set of $N$ point-scatterers shown in Fig.~\ref{fig1}. Assume that scatterer \#1 is excited by a unit field, and follow the wave-field that hits and excites scatterer \#2, then \#3 etc..., until it completes a closed loop.  Since all scatterers are electrically small, their excitation can be faithfully described by the associated polarizabilities $\alpha_n, n=1,\ldots N$. It means that the excitation of the $n$-th scatterer is given by $\alpha_nE^L(\rhov_n)$, where $E^L(\rhov_n)$ is the local field at the $n$-th scatterer center: the field there, in the absence of that specific scatterer. Hence, at the completion of the loop the field is given by,
\begin{equation}\label{eq7}
E = \Pi_{m=1}^N\alpha_m G(\rhov_{m},\rhov_{m-1})
\end{equation}
where we set $\rhov_0\equiv\rhov_N$. With \Eq{eq6}, the last result can be re-written as
\begin{subequations}
\begin{equation}\label{eq8a}
E=\left[\Pi_{m=1}^N\alpha_m\Gst(\rhov_{m},\rhov_{m-1})\right]e^{i\phi}
\end{equation}
The expression in the square brackets is the field after it accomplishes a complete loop in a $\mathcal{R}^0$ problem, which inclues the effect of The refraction index $n$. Since the footprint of rotation on the scatterers' polarizabilities is negligible \cite{KazmaSteinbergEMTS2019}, it manifests only in $\phi$, given by,
\begin{equation}\label{eq8b}
\phi = 2k_0\frac{\Omega}{c}\sum_{m=1}^NS_{\triangle_m}=
2k_0\frac{\Omega}{c}S
\end{equation}
\end{subequations}
where $S_{\triangle_m}=\unit{z}\cdot(\rhov_{m-1}\times\rhov_{m})/2$ is the \emph{area} of the triangle whose vertices are the origin (=rotation axis). $S\equiv \sum_mS_{\triangle_m}$ is the total area enclosed by the loop, and it follows that $\phi$ is \emph{independent} of the refraction index $n$ of the surrounding homogeneous medium. This result is nothing but the celebrated Sagnac phase-shift \cite{Post}. Also, we note that the total area enclosed by the loop becomes negative if the series of events is inverted (representing propagation in the opposite direction). This fact establishes the well known Sagnac phase shift $\Delta\phi=4k_0\Omega|S|/c$ between clockwise and counter clockwise propagating waves.
Finally, it is straightforward to show that these results hold also if the rotation axis (=the origin) is outside of the closed loop.

The purpose of the analysis above is to verify that the Sagnac phase shift is correctly preserved in the approximation; this is important since it enables us to use a simple and physically transparent expression for $G$ (and consequently for the phases) in subsequent derivations. Furthermore, the new results discussed below seem to contradict Sagnac effect, but this apparent contradiction is derived by the very same Green's function that correctly reconstructs Sagnac effect, and therefore they are mathematically consistent.

\begin{figure}[ht]
\centering\includegraphics[scale=0.28]{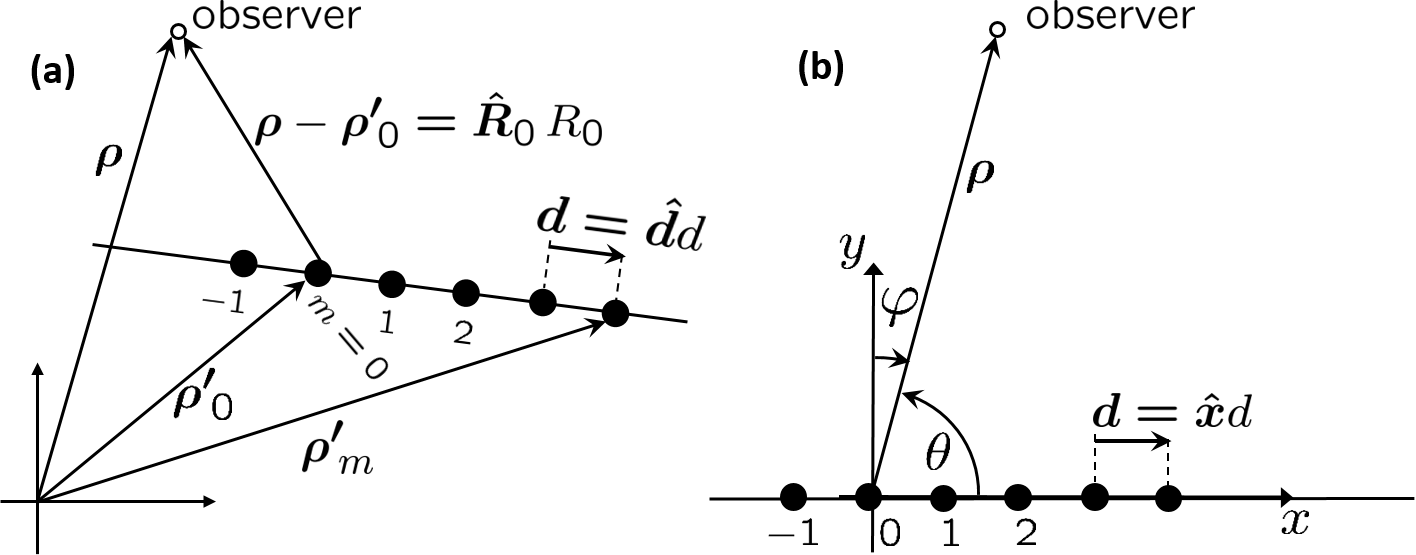}
\caption{A periodic array of point-scatterers. (a) The general setting. (b) The effect of rotation is maximized in regions where $\rhov\perp \unit{d}$.}
\label{fig2}
\end{figure}

\noindent
{\bf \emph{A paradigm interference experiment}}
Consider an infinite linear periodic array of point scatterers shown in Fig.~\ref{fig2}, observed in $\mathcal{R}^\Omega$. For simplicity, we assume that the excitation magnitude of all scatterers is the same, and the interference pattern of the system is observed in the far-field. Since the scatterers are electrically small the radiation of the $m$-th scatterer is given by $G(\rhov,\rhovp_{\!\! m})$ where
$\rhovp_{\!\! m}$ is its location,
\begin{equation}\label{eq9}
\rhovp_{\!\! m}=\rhovp_{\!\! 0}+m\bb{d}=\rhovp_{\!\! 0}+md\unit{d},
\end{equation}
$\bb{d}=d\unit{d}$ is the array lattice vector, and $d$ is the period length. The interference field at an arbitrary observation point $\rhov$ is given by the Green's function summation [use \Eq{eq6}]
\begin{subequations}
\begin{eqnarray}
E(\rhov)&=&\sum_m G(\rhov,\rhovp_{\!\! m})\label{eq10}\\
 &=& \sum_m \Gst(\rhov,\rhovp_{\!\! m})e^{ik_0\frac{\Omega}{c}\,\unit{z}\cdot(\rhovp_{\!\! m}\times\rhov)}.\label{eq10b}
\end{eqnarray}
\end{subequations}

We invoke now the standard algebraic procedure commonly used to obtain the far-field. First, use the large argument approximation of the Hankel function in $\Gst$: $H_0^{(1)}(x)\simeq\sqrt{2/(\pi x)}e^{i(x-\pi/4)}$. In the amplitude we approximate $1/\abs{\rhov-\rhovp_{\!\! m}}\simeq 1/R_0\,\,\forall\,m$, where $R_0=\abs{\rhov-\rhovp_{\!\! 0}}$. In the exponent we set
$\abs{\rhov-\rhovp_{\!\! m}}\simeq R_0-m\bb{d}\cdot\unit{R}_0$ where $\unit{R}_0$ is a unit vector defined by $\unit{R}_0=(\rhov-\rhovp_{\!\! 0})/R_0$. Substituting it back to \Eq{eq10} we get with no further approximation
\begin{subequations}
\begin{equation}\label{eq11a}
E\simeq\frac{e^{i\pi/4}e^{ik_0\left[nR_0-\frac{\Omega}{c}\unit{z}\cdot(\rhov\times\rhovp_{\!\! 0})\right]}}{(8\pi k_0nR_0)^{1/2}}\,
\sum_{m'}e^{-im'q},
\end{equation}
where
\begin{equation}\label{eq11b}
q=k_0d\,\left[n\unit{d}\cdot\unit{R}_0+\frac{\Omega}{c}\unit{z}\cdot(\rhov\times\unit{d}\,)\right].
\end{equation}
\end{subequations}
Clearly, the field in \Eq{eq11a} peaks when $q$ becomes an integer multiple of $2\pi$. Therefore the IM are obtained at $\rhov=\rhov_\mathfrak{M}$ given by ($\mathfrak{M}=0,\pm1,\pm2,\ldots $)
\begin{equation}\label{eq12}
n\unit{d}\cdot\unit{R}_0^\mathfrak{M} + \frac{\Omega}{c}\unit{z}\cdot(\rhov_\mathfrak{M}\times\unit{d})=\mathfrak{M}\frac{\lambda}{d},
\end{equation}
where $\unit{R}_0^\mathfrak{M}\equiv (\rhov_\mathfrak{M}-\rhovp_{\!\! 0})/\abs{\rhov_\mathfrak{M}-\rhovp_{\!\! 0}}$, and $\lambda$ is the vacuum wavelength in $\mathcal{R}^0$. The rotation footprint is most profound where $\rhov\perp\unit{d}$. In these regions the specific choice of $\rhov_0$ has no significant role, so one may conveniently set $\rhov_0=\bb{0}$. This specific setting is illustrated in Fig.~\ref{fig2}(b). Assume now that the interference is observed on a screen parallel to the array axis, located at a constant range $y>0$. We define $\varphi_\mathfrak{M}=\pi/2-\theta_\mathfrak{M}$ as the angle between the $y$ axis and $\rhov_\mathfrak{M}$ ($-\pi/2<\varphi_\mathfrak{M}<\pi/2$). Then, for the specific setting in Fig.~\ref{fig2}(b) the general expression in \Eq{eq12} yields (see appendix \ref{eq9der})
\begin{equation}\label{eq13}
\varphi_\mathfrak{M}=\arcsin\left[\frac{1}{n}\left(\frac{\mathfrak{M}\lambda}{d}+\frac{\Omega y}{c}\right)\right].
\end{equation}
For $\Omega=0$ it boils down to the traditional result of diffraction grating in $\mathcal{R}^0$. It is seen that rotation adds a \emph{range-dependent} term, making the IM follow \emph{curved trajectories} in space with bending that depends on the background refraction index $n$. This is to contrast with the classical Sagnac effect, reconstructed above, that is \emph{medium independent}. The bending scales as $n^{-1}$, and it exists already at the zeroth-order diffraction term ($\mathfrak{M}=0$). Therefor low-index increases sensitivity to rotation.
this enhancement, however, is not unbounded since $n$ itself has a lower bound: $n>\frac{\mathfrak{M}\lambda}{d}+\frac{\Omega y}{c}$ to render $\varphi_\mathfrak{M}$ real. For $\mathfrak{M}=0$ it yields $n>\Omega y/c$. For example assume $\Omega\approx 7.3\times 10^{-5} \mbox{ sec}^{-1}$ (earth rotation). For $y=1$m, $n>2.4\times 10^{-13}$. Hence one can get enhancement of many orders of magnitude before the lower bound on $n$ is approached. As $\Omega$ gets smaller, so does the lower bound on $n$, thus enabling a stronger enhancement of sensitivity to rotation.

\begin{figure}[ht]
\centering\includegraphics[scale=0.31]{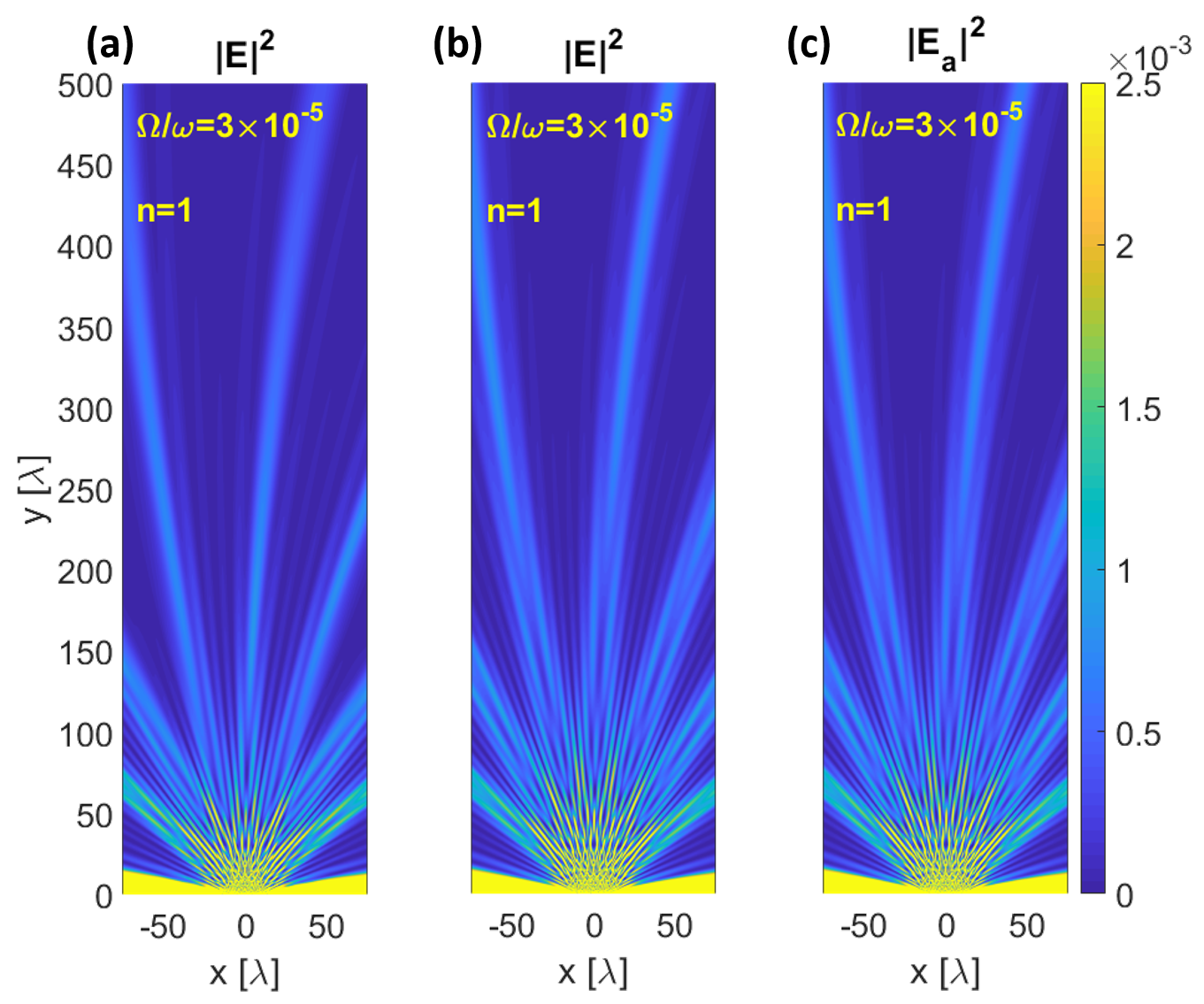}
\caption{Interference patterns due to eight point-scatterers. (a) Using \Eq{eq10} with the formal expansion in \Eq{Aeq5a} for $G$ with 161 cylindrical harmonics. (b) With 321 cylindrical harmonics. (c) Using the Green's function in \Eq{eq6}. The coordinates are normalized to the vacuum wavelength.  The axes and colormaps are the same for all sub-figures. The IM match \Eq{eq13}.}
\label{fig3}
\end{figure}

We computed the interference field by using \Eq{eq10} with the exact Green's function in \Eq{Aeq5a} ($G_{exact}$), and using the simplified expression in \Eq{eq6} above. The system consists of eight points located at $y_m=0$ and $x_m=\lambda[-14,-10,-6,-2,2,6,10,14]$, where $\lambda$ is the vacuum wavelength, and the background is vacuum. The results are shown in Figs.~\ref{fig3}(a)-(c) for $\Omega/\omega=3\times 10^{-5}$. The $x,y$ coordinates are in units of $\lambda$. Figure \ref{fig3}(a) shows the interference as obtained from \Eq{eq10} with $G_{exact}$ with $\ell={-80,...,80}$. Figure \ref{fig3}(b) shows the same but with $\ell={-160,...,160}$. It is seen that the patterns carry the same physical picture of curved IM, but they differ in some details. Figure \ref{fig3}(c) shows the interference computed with the approximate Green's function in \Eq{eq10b}. It is identical to Fig.~\ref{fig3}(b).
In all cases, the IM follow the curved trajectories predicted by \Eq{eq13}.

\begin{figure}[ht]
\centering\includegraphics[scale=0.3]{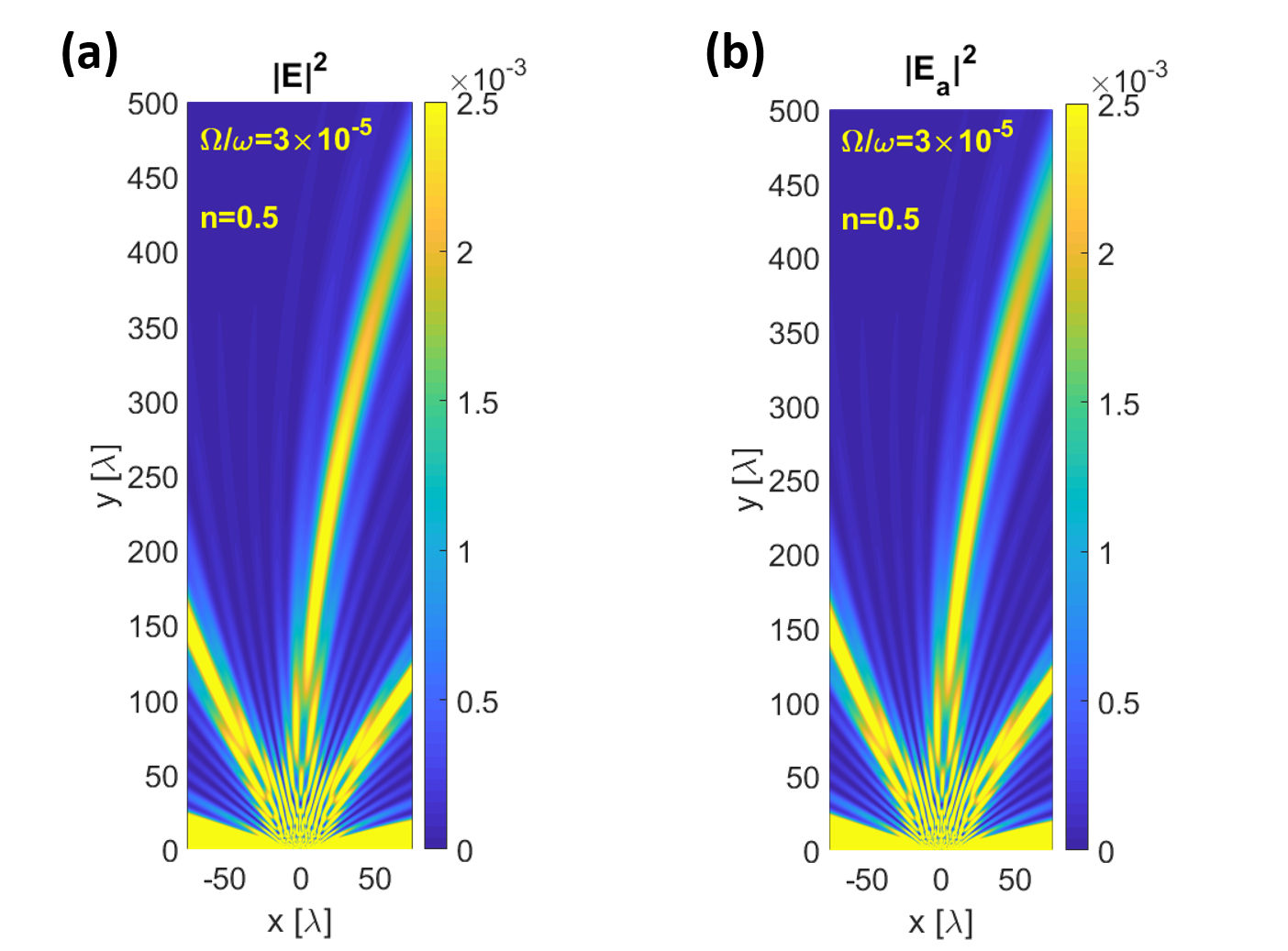}
\caption{The same as in Fig.~\ref{fig3}, but with a background medium with refraction index $n=0.5$. (a) With $G$ in \Eq{Aeq5a} and 321 cylindrical harmonics. (b) Using $G$ of \Eq{eq6}. Bending of the IM increased according to \Eq{eq13}.}
\label{fig4}
\end{figure}

Figure \ref{fig4} shows the interference of the same system as in Fig.~\ref{fig3}, but now the background refraction index is $n=0.5$. The IM bending increased significantly, and matches \Eq{eq13}. This last observation is counter intuitive. Lower $n$ implies faster wave, and one would expect a faster wave to bend \emph{less}, while \Eq{eq13} predicts the contrary. We clarify this apparent contradiction below.

Finally, we note that the Fresnel drag coefficient (or Fizeau drag) \cite{LP1} $\alpha=1-n^{-1}$ measures the degree of non-reciprocity of light propagation in moving medium in the inertial frame, and it has been related to the space-time metric. Low $n$ may invert the drag direction, and even the metric sign. This intriguing observation needs to be clarified by further study.

\noindent
{\bf \emph{Energy and momentum pictures}}
The time-averaged power flux and momentum densities associated with the wave \emph{only}, in non-inertial frame, are still given by $\bb{S}=\frac{1}{2}\Re(\bb{E}\times\bb{H}^*)$ and by $\bb{S}/c^2$, respectively \cite{AM_Ramos_2015}. The latter is nothing but the Abraham momentum picture \cite{AM_Ramos_2015, AM_Barnett_2010}. $\bb{S}$ for a line source of electric current is given by \Eq{Aeq8a} - \Eq{Aeq8e}. For $\rho'/\rho\rightarrow 0$ we have $S_\rho\propto n$ and $\Omega$-independent, while $S_\theta\propto\Omega$ and has $n$-independent magnitude.
Thus the \emph{ratio} $S_\theta/S_\rho$ increases as $n$ decreases, which results in stronger bending. We note that in the Minkowski picture the momentum density is $\frac{1}{2}\Re(\bb{P}=\bb{D}\times\bb{B}^*)$. This expression, however, does not distinguish between the wave and medium parts and therefore it is less convenient for exposing the mechanism underlying the effect of $n$ on the field interference bending.

Since the curved IM in \Eq{eq13} and Figs.~(\ref{fig3})-(\ref{fig4}) are due to interference, a better description that is consistent with the Minkowski picture \emph{and} takes into account the medium effect, is based on a \emph{quasi-particle} approach. Here, the rotating medium effect enters through a Coriolis force $\bb{F}_c$ acting on the quasi-particle, resulting in curved trajectory. However, since the Sagnac effect is independent of the medium's refraction index $n$ \cite{Post} and so is the rotation induced phase in our Green's function analysis, the \emph{consistent} model of $\bb{F}_c$ must also be $n$-independent. Therefore we express it as $F_c=2\bb{p}_0\times\bb{\Omega}$ where $\bb{p}_0$ is the \emph{free space} momentum. This force is identical to the rate of change of the quasi-particle actual momentum $\bb{p}$, that according to Minkowski model is $\bb{p}=\bb{p}_0n$. We arrive at the equation of motion
\begin{equation}\label{eq14}
2\bb{p}_0\times\bb{\Omega}=\dot{\bb{p}}_0 n
\end{equation}
where the over-dot indicates a time derivative. Hence $\dot{\bb{p}}_0=n^{-1}2\bb{p}_0\times\bb{\Omega}$. This directly implies that the curvature of the quasi-particle trajectory is inversely proportional to $n$. Thus lower index induces tighter bending.

\noindent
{\bf \emph{Summary} }
A concise theory of radiation from point sources embedded in a rotating medium, and observed in its rest-frame, was presented. The formulas developed are easy to employ in any physical system. The formulation was used to illustrate the Sagnac effect as a simplified series of nearest-neighbor scattering events, leading to the known result where the  rotation induced phase depends on the area enclosed by the scatterer array, and is medium \emph{independent}. This formulation was then used to treat the problem of diffraction from a linear array of emitters embedded in a rotating medium, giving rise to rotation induced ``bending'' of the interference pattern. Unlike the Sagnac effect, the bending is medium \emph{dependent} with a curvature inversely proportional to the rotating medium refractive index. Therefore, low-index metamaterials can be used to enhance rotation-induced effects. This result is counter intuitive as one expects that a faster wave would experience weaker bending, and lends itself to interpretations in a dual way.
From the calculated fields point of view, the bending is associated with the inverse dependence of the power flux and Abraham momentum on the refractive index. From a quasiparticle-diffraction point of view, we can treat the bending of the fields using the action of an equivalent Coriolis type force, where the Minkowski momentum picture is the appropriate one here, in a manner consistent with the interpretation presented in \cite{AM_Barnett_2010}.

\vspace*{0.1in}
\noindent
{\bf \emph{Acknowledgement} }
The author BZS gratefully acknowledges fruitful discussions
with Prof. Hui Cao at Yale.

\newpage
\begin{widetext}
\appendix
\section{Introduction and theoretical considerations}\label{AppIntro}
We present a preliminary study of the effect of slow rotation on the interference and diffraction patterns generated by a set of electrically small scatterers. Our analysis is two-dimensional; that is, the scatterers geometry and the electromagnetic excitation are invariant along $\unit{z}$ - the medium's rotation axis. The entire structure rotates at a constant angular velocity $\bb{\Omega}=\unit{z}\Omega$ radians/Sec., and it is observed at its rest frame of reference $\mathcal{R}^{\Omega}$. Here and henceforth vectors are written in bold letters, and a hat indicates a unit-vector. In the \emph{rotating} frame of reference $\mathcal{R}^{\Omega}$ the coordinates normal to the rotation axis are denoted by $\bb{\rho}=\unit{x}x+\unit{y}y=\rho\unit{\rho}$.
We limit our study to slowly rotating structures, meaning that $\Omega\rho_{\mbox{max}}/c\ll 1$ where ${\mbox{max}}$ is the maximal distance of the system (including possible sources or observation points) from the rotation axis. The system is said to be \emph{stationary} in the special case of $\Omega=0$. In the latter case, the reference frame $\mathcal{R}^\Omega=\mathcal{R}^0$ is inertial.
We use the term ``a problem in $\mathcal{R}^{\Omega}$'' to designate the electrodynamic problem of a rigid structure that rotates at an angular velocity $\Omega$, and observed in its \emph{non-inertial} rest frame. A specific case of which, ``a problem in $\mathcal{R}^{0}$'' is the electrodynamic problem of a rigid structure \emph{at rest}, observed in the inertial Lab frame. We always observe the rigid structure in its rest frame of reference.

\subsection{Summary of previous results}

We assume that the materials in the corresponding stationary system are fully characterized by scalar permittivity $\epsilon(\bb{\rho})$ and permeability $\mu(\bb{\rho})$. The electrodynamics of a slowly rotating medium, as observed in the medium's rest-frame, is governed by the same set of Maxwell equations as for the stationary case except that the constitutive relations are modified to account for the rotation effect \cite{Shiozawa} (a time-harmonic dependence $e^{-i\omega t}$ is assumed and suppressed)
\begin{subequations}
\begin{eqnarray}\label{Aeq1}
\nabla\times\bb{E} &=& i\omega\bb{B},\qquad \nabla\cdot\bb{B}=0\label{Aeq1a}\\
\nabla\times\bb{H} &=& -i\omega\bb{D},\qquad \nabla\cdot\bb{D}=0\label{Aeq1b}
\end{eqnarray}
\end{subequations}
where the rotation is manifested in the constitutive relations
\begin{equation}\label{Aeq2}
\bb{B}=\mu\bb{H}+c^{-2}\left(\bb{\Omega}\times\rv\right)\times\bb{E},\quad
\bb{D}=\epsilon\bb{E}-c^{-2}\left(\bb{\Omega}\times\rv\right)\times\bb{H}.
\end{equation}
In the above, $c$ is the speed of light in vacuum, as observed in $\mathcal{R}^0$ (i.e.~in an inertial frame of reference). Likewise, $\mu=\mu_0\mu_r$, $\epsilon=\epsilon_0\epsilon_r$, and $\mu_r,\epsilon_r$ are the material properties as observed in $\mathcal{R}^0$.
A derivation of these constitutive relations directly from a more general form of Maxwell's equations can be found e.g.~in \cite{AndersonRyon}

Our starting point is the Green's function theory developed in \cite{BZSG}. It has been shown that in 2D problems the electromagnetic field can be rigorously separated into completely decoupled TE and TM polarizations:

\begin{description}
\item[TE] This case consists of the fields $H_z,\bb{E}_t=\unit{x}E_x+\unit{y}E_y$ only. These fields can be excited by the transverse electric current $\bb{J}_t$ and/or by the $\unit{z}$-directed magnetic current $J_z^M$.

 \item[TM] This case consists of the fields $E_z,\bb{H}_t=\unit{x}H_x+\unit{y}H_y$ only. These fields can be excited by the transverse magnetic current $\bb{J}^M_t$ and/or by the $\unit{z}$-directed electric current $J_z$.
     \end{description}
Furthermore, in both polarizations the complete electromagnetic field can be derived from the corresponding $z$-directed field, that satisfies the following transverse PDE
\begin{subequations}
\begin{equation}\label{Aeq3a}
[\nabla_t^2+k_0^2n^2]F_z-2ik_0^2\frac{\Omega}{\omega}\partial_\theta\,F_z=S
\end{equation}
where $n^2=\epsilon_r\mu_r$. For TE polarization the complete EM field is given by,
\begin{equation}\label{Aeq3b}
H_z=F_z,\quad i\omega\epsilon\bb{E}_t=\bb{J}_t-\nabla_t\times\unit{z}H_z+i\frac{\omega\Omega}{c^2}\bb{\rho}H_z
\end{equation}
with the scalar source $S$
\begin{equation}\label{Aeq3c}
S=S^{\mbox{\tiny TE}}=-i\omega\epsilon J_z^M-i\frac{\omega\Omega}{c^2}\bb{\rho}\cdot\bb{J}_t-\unit{z}\cdot\nabla_t\times\bb{J}_t
\end{equation}
For TM polarization
\begin{equation}\label{Aeq3d}
E_z=F_z,\quad i\omega\mu\bb{H}_t=\bb{J}_t^M+\nabla_t\times\unit{z}E_z-i\frac{\omega\Omega}{c^2}\bb{\rho}E_z
\end{equation}
with the scalar source
\begin{equation}\label{Aeq3e}
S=S^{\mbox{\tiny TM}}=-i\omega\mu J_z+i\frac{\omega\Omega}{c^2}\bb{\rho}\cdot\bb{J}_t^M+\unit{z}\cdot\nabla_t\times\bb{J}_t^M
\end{equation}
\end{subequations}

We define now the scalar Green's function $G(\bb{\rho},\bb{\rho'})$ as the impulse response of the PDE in \Eq{Aeq3a},
\begin{equation}\label{Aeq4}
[\nabla_t^2+k_0^2n^2]G-2ik_0^2\frac{\Omega}{\omega}\partial_\theta\,G=-\delta(\bb{\rho}-\bb{\rho'})=
-\frac{1}{\rho'}\delta(\rho-\rho')\delta(\theta-\theta')
\end{equation}
Clearly, the response to any excitation can be obtained by a simple spatial convolution of $G$ with the source term $S$. Note, however, that for a thin-wire electric [magnetic] current $\bb{J}=\unit{z}I_z\delta(\bb{\rho}-\bb{\rho'})$ [$\bb{J}^M=\unit{z}I_z^M\delta(\bb{\rho}-\bb{\rho'})$], the excited field is only of the TM [TE] polarization, with the Electric [magnetic] field $E_z=i\omega\mu I_z\,G(\bb{\rho},\bb{\rho'})$ [$H_z=i\omega\epsilon I^M_z\,G(\bb{\rho},\bb{\rho'})$].

It has been shown in \cite{BZSG} that a formally complete and \emph{exact} expression for $G$ is given by
\begin{subequations}
\begin{equation}\label{Aeq5a}
G(\rhov,\rhovp)=\frac{i}{4}\sum_{\ell=-\infty}^\infty J_\ell(k_0 n \gamma_\ell\rho_<)H_\ell^{(1)}(k_0 n \gamma_\ell\rho_>)e^{i\ell(\theta-\theta')}
\end{equation}
where
\begin{equation}\label{Aeq5b}
\gamma_\ell=\sqrt{1+2\ell\Omega/(\omega n^2)},\qquad \rho_\gtrless=\substack{\mbox{max} \\ \mbox{min}}(\rho,\rho').
\end{equation}
\end{subequations}
There are several difficulties with the summation in \Eq{Aeq5a}. First and foremost, recall that $G(\rhov,\rhovp)$ should diverge logarithmically as $\rhov\rightarrow\rhovp$ (as any Greens function of a 2D PDE). This singular behavior is not present in each of the summed terms above (unless $\rho=\rho'=0$). Thus, as $\rhov$ approaches $\rhovp$ the convergence of the series in \Eq{Aeq5a} becomes extremely poor. The second difficulty stems from the fact that we will be interested in far-field diffraction/interference patterns. This requires the evaluation of $J_\ell,\, H_\ell^{(1)}$ for large argument and order, that is also quite difficult to attain. Finally, while the cylindrical harmonic summation in \Eq{Aeq5a} provides a formally exact Green's function of the rotating medium, the physically meaningful and intuitive entities such as amplitude, phase, optical path, etc..., are completely hidden.
In light of these issues, an approximate expression for $G$ has been developed in \cite{BZSG} with the condition $\max(\rho,\rho')\Omega/c\ll1$ that is consistent with the slow rotation assumption,
\begin{equation}\label{Aeq6}
G(\rhov,\rhovp)\simeq\Gst(\rhov,\rhovp)e^{ik_0(\Omega/c)\unit{z}\cdot(\rhovp\times\rhov)},\quad
\Gst(\rhov,\rhovp)=\frac{i}{4}H_0^{(1)}(k_0n|\rhov-\rhovp|)
\end{equation}
where $H_0^{(1)}(\cdot)$ is the zeroth-order Hankel function of the first kind. Clearly, $\Gst$ is the Green's function of the corresponding \emph{stationary} medium; the homogeneous medium Greens function for a problem in $\mathcal{R}^0$. Note that the singularity of $G$ as $\rhov\rightarrow\rhovp$ is fully encapsulated in $\Gst$, since the other term is regular in this limit. This is quite expected since both the stationary and rotating medium Green's functions satisfy a PDE that possesses exactly the same form of the higher (second-order) derivatives - see \Eq{Aeq4} - and therefor must have the same singularity; the stationary case is obtained merely by setting $\Omega=0$. Hence, the simplified expression above \emph{correctly reconstructs the near fields}. Despite its simplicity, this expression provides a good approximation for the rotating medium Green's function. A detailed comparison is performed in \cite{BZSG}. We show in the main text that this Green's function also correctly reconstructs the celebrated Sagnac effect that is clearly related to propagating fields. Hence it constitutes in fact a \emph{uniform} approximation of the exact Green's function.

\section{Derivation of \Eq{eq13}}\label{eq9der}

If we set $\bb{\rho}_0=\bb{0}$ then Eq.~(8) in the main text becomes
\begin{equation}
n\unit{x}\cdot\unit{\rho}_\mathfrak{M}+\frac{\Omega}{c}\unit{z}\cdot\left(\bb{\rho}_\mathfrak{M}\times\unit{x}\right)=\mathfrak{M}\lambda/d\label{eqR1}
\end{equation}
where $\unit{\rho}_\mathfrak{M}=\unit{x}\sin\varphi_\mathfrak{M}+\unit{y}\cos\varphi_\mathfrak{M}$ is the unit vector pointing towards the interference $\mathfrak{M}$-th maxima, $\varphi_\mathfrak{M}$ is the value of the angle $\varphi$, as defined in Fig.~2(b) of the main text, that points to that maxima, and $\bb{\rho}_\mathfrak{M}=\rho\unit{\rho}_\mathfrak{M}$. By substituting this back to \Eq{eqR1} we obtain
\begin{equation}
n\sin\varphi_\mathfrak{M}=\mathfrak{M}\lambda/d+\Omega y/c \label{eqR2}
\end{equation}
where we used $y=\rho\cos\varphi_\mathfrak{M}$. This is the result shown in the main text.

\section{Fields in $\mathcal{R}^\Omega$}

Assume a $\unit{z}$-directed electric line source $I_z$ located at $\rhovp=(\rho',\theta')$ in $\mathcal{R}^\Omega$. It excites only the TM polarization. The corresponding electric and magnetic fields can be obtained from $G$ via \Eqs{Aeq3d}{Aeq3e},
\begin{subequations}
\begin{eqnarray}
E_z &=& I_z i\omega\mu G(\rhov,\rhovp)\label{Aeq7a}\\
\nonumber\\
H_\rho &=& I_z \frac{k_0 n\rho'\sin(\theta-\theta')}{\abs{\rhov-\rhovp}}\Gstp
e^{ik_0(\Omega/c)\unit{z}\cdot(\rhovp\times\rhov)} \nonumber\\
 & & -I_z i\frac{k_0\Omega}{c}
\left[\rho-\rho'\cos(\theta-\theta')\right]G(\rhov,\rhovp)\label{Aeq7b}\\
\nonumber\\
H_\theta &=& -I_z\frac{k_0 n[\rho-\rho'\cos(\theta-\theta')]}{\abs{\rhov-\rhovp}}\Gstp e^{ik_0(\Omega/c)\unit{z}\cdot(\rhovp\times\rhov)}\nonumber\\
 & & -I_z i \frac{k_0\Omega}{c}\rho'\sin(\theta-\theta')G(\rhov,\rhovp) \label{Aeq7c}
\end{eqnarray}
where for simplicity and convenience, we have used \Eq{eq6} in the derivation of the magnetic field. In the expressions above, $\Gstp$ is the derivative of $\Gst$ with respect to the argument,
\begin{equation}\label{Aeq7d}
\Gstp=-\frac{i}{4}H_1^{(1)}(k_0n\abs{\rhov-\rhovp}).
\end{equation}
\end{subequations}
Recall that Maxwell equations for a problem in $\mathcal{R}^\Omega$ differ from their form for a problem in $\mathcal{R}^0$ only via the more elaborated form of the constitutive relations, given here in \Eq{Aeq2}. We note that these constitutive relations may hold also for a problem in $\mathcal{R}^0$ with a special type of non-homogeneous Telegen medium \cite{Bian_Book}. The time-averaged power flux and momentum densities associated with the wave \emph{only}, in non-inertial frame or in a general medium with macroscopic constitutive relations of the type dealt with here, are still given by $\bb{S}=\frac{1}{2}\Re(\bb{E}\times\bb{H}^*)$ and by $\bb{S}/c^2$, respectively \cite{AM_Ramos_2015}. The latter is nothing but the Abraham momentum picture \cite{AM_Ramos_2015, AM_Barnett_2010}. We have,
\begin{subequations}
\begin{eqnarray}
S_\theta &=& S_\theta^0 -4P_0\frac{k_0\Omega}{c}[\rho-\rho'\cos(\theta-\theta')]\abs{\Gst}^2 \label{Aeq8a}\\
S_\rho &=& S_\rho^0 + 4P_0\frac{k_0\Omega}{c}\rho'\sin(\theta-\theta')\abs{\Gst}^2\label{Aeq8b}
\end{eqnarray}
where $S_\theta^0$ and $S_\rho^0$ are rotation independent terms that provide the $\unit{\theta}$ and $\unit{\rho}$ components of the Poynting vector for a problem in $\mathcal{R}^0$,
\begin{eqnarray}
S_\theta^0 &=& 4P_0 k_0n\frac{\rho'\sin(\theta-\theta')}{\abs{\rhov-\rhovp}}\,\Re\left\{i\Gst\left[\Gstp\right]^*\right\}\label{Aeq8c}\\
S_\rho^0 &=& 4P_0 k_0n\frac{\rho-\rho'\cos(\theta-\theta')}{\abs{\rhov-\rhovp}}\,\Re\left\{i\Gst\left[\Gstp\right]^*\right\}\label{Aeq8d}
\end{eqnarray}
and where
\begin{equation}\label{Aeq8e}
P_0=\frac{1}{8}\omega\mu\abs{I_z}^2
\end{equation}
\end{subequations}
is the time-averaged power per unit length radiated by a wire carrying the electric current $I_z$ in a problem in $\mathcal{R}^0$.
It follows that low index material can reduce significantly the rotation independent terms $S_\theta^0$ and $S_\rho^0$, but has no effect at all on the rotation-dependent components of $\bb{S}$.

The excitation of TE polarization, e.g. by a $\bb{z}$-directed line-source of magnetic current, has a symmetric structure and will not be repeated here.

\end{widetext}

%
%
%
%

%
%
\end{document}